\documentclass[preprint,aps,amsmath,amssymb,nofootinbib,tightenlines,superscriptaddress]{revtex4}
\usepackage{amsmath,amssymb}

\usepackage{hyperref}
\usepackage{graphicx}
\usepackage{slashed}

\newcommand{\phref}[1]{\href{http://arxiv.org/abs/#1}{arXiv:#1 [hep-ph]}}

\begin{document}

\vspace{3.0cm}
\preprint{\vbox 
{
\hbox{WSU--HEP--1305} 
}}

\vspace*{2cm}

\title{Nonstandard neutrino interactions and transition magnetic moments}

\author{Kristopher J. Healey}%
\email{healey@wayne.edu}%
\affiliation{Department of Physics and Astronomy, Wayne State University,
Detroit, Michigan 48201, USA}

\author{Alexey A. Petrov}%
\email{apetrov@wayne.edu}%
\affiliation{Department of Physics and Astronomy, Wayne State University,
Detroit, Michigan 48201, USA}%
\affiliation{Michigan Center for Theoretical Physics, University of Michigan,
Ann Arbor, Michigan 48109, USA}

\author{Dmitry Zhuridov}%
\email{dmitry.zhuridov@wayne.edu}%
\affiliation{Department of Physics and Astronomy, Wayne State University,
Detroit, Michigan 48201, USA}

\date{\today \\ \vspace{1in}}

\begin{abstract}
We constrain generic nonstandard neutrino interactions with existing experimental data on neutrino transition magnetic moments and derive strong bounds on tensorial couplings of neutrinos to charged fermions. We also discuss how some of these tensorial couplings can be constrained by other experiments, e.g., on neutrino-electron and neutrino-nucleus scattering.
\end{abstract}


\maketitle


Neutrinos have long been a prime vehicle for testing the standard model (SM) of particle interactions. They
played instrumental role in measurements of parton distribution functions~\cite{Peskin:1995ev}, quark mixing parameters~\cite{Artuso:2008vf} 
and other important quantities. Neutrino oscillations provided the first glimpse of the physics beyond the minimal standard 
model by establishing that neutrinos have mass~\cite{PDG2012}. It would not be entirely surprising if other signs of new physics 
were revealed in the precision studies of neutrino properties. It is therefore important to study deviations
of neutrino interaction parameters with other SM particles from their SM expectations. There have been many analyses of 
nonstandard neutrino interactions, often abbreviated as NSIs, in neutrino scattering and oscillation 
experiments~\cite{Lee:1977tib,Grossman:1995wx,Bergmann:1999rz,Barranco:2011wx,Kopp:2007ne,Ohlsson:2012kf,Rashed:2013dba}. They have been 
particularly important in the studies leading up to a possible future neutrino factory. In this paper we 
point out that it is possible to constrain NSIs using existing measurements of neutrino transition magnetic moments. 

It has been seen that nonstandard  neutrino interactions play a subdominant role in neutrino scattering. 
Provided that the scale of new physics $M$ is large compared to the electroweak scale, the 
easiest parameterization of NSIs of $\nu\nu ff$ type at low-energy scales accessible in neutrino experiments would naturally be
written in terms of effective four-fermion operators of dimension 6~\cite{Bergmann:1999rz,Kopp:2007ne,Kingsley:1974kq,Cho:1976um,Kayser:1979mj},
\begin{eqnarray}\label{EffLgr}
-\mathcal{L}_\text{eff}= \sum_a
\frac{\epsilon_{\alpha\beta}^{fa}}{M^2} (\bar\nu_\beta \Gamma_a \nu_\alpha) (\bar f \Gamma_a f) 															+	{\rm h.c.},
\end{eqnarray}
where $\epsilon_{\alpha\beta}^{fa}$ are NSI couplings, $f$ denotes the component of an arbitrary weak doublet (often an electron or a quark field for 
studies of $\nu$NSIs in matter), $\Gamma_a=\{I,\gamma_5,\gamma_\mu,\gamma_\mu \gamma_5,\sigma_{\mu\nu}\}$, 
$a=\{S,P,V,A,T\}$ and $\sigma_{\mu\nu} = i[\gamma_\mu,\gamma_\nu]/2$. Effective nonstandard neutrino interactions, 
expressed by the four-fermion operators as in Eq.~(\ref{EffLgr}), are widely discussed in the literature, 
see Refs.~\cite{Kopp:2007ne,Ohlsson:2012kf,Rashed:2013dba} for recent reviews. 
Typically only left-handed neutrinos are considered, which allows the study of NSIs' impact on solar, atmospheric, and reactor neutrinos, 
as well as on neutrino-nucleus scattering. Also, oftentimes, only left-handed or right-handed {\it vectorial} interactions are 
considered.

It is important to note that this restriction removes from the consideration a large class of models in which neutrino interactions could violate
lepton number, e.g., models with leptoquarks and R-parity-violating supersymmetric theories. The effective low-energy operators  that are 
generated in those models include
\begin{eqnarray}\label{EffLgr2}
-\mathcal{L}_\text{eff} \subset  \sum_a
\frac{\tilde\epsilon_{\alpha\beta}^{fa}}{M^2} (\bar\nu_\beta \Gamma_a f) (\bar f \Gamma_a  \nu_\alpha) 															+	{\rm h.c.}
\end{eqnarray}
Using Fierz identities, Eq.~(\ref{EffLgr2}) can be rewritten in the form of Eq.~(\ref{EffLgr}) if all effective operators, 
including the tensor ones, are considered. It is therefore important to consider an effective Lagrangian that also includes
tensorial interactions. The chirality constraint that allows $\nu\nu ff$ interaction only of $V\pm A$ types cannot describe 
possible important neutrino phenomena, such as neutrino magnetic moment (NMM).  It is the tensor interactions 
of neutrinos that we will attempt to constrain in this paper.

Neutrino magnetic moment $\mu_{\alpha\beta}$ can be defined by the Hermitian form factor 
${f^M_{\alpha\beta}(0) \equiv \mu_{\alpha\beta}}$ of the term~\cite{Broggini:2012df}
\begin{eqnarray}
	-f^M_{\alpha\beta}(q^2)~   \bar\nu_\beta(p_2) \, i\sigma_{\mu\nu} q^\nu  \nu_\alpha(p_1)
\end{eqnarray}
in the 
effective neutrino electromagnetic current,
where $\alpha,\beta=e,\mu,\tau$ are flavor indices, $q=p_2-p_1$. 
The relation between NMMs in the flavor basis and in the mass basis can be written as~\cite{Broggini:2012df,Giunti:2008ve,Beacom:1999wx}
\begin{eqnarray}\label{eq:flavorNMM}
		\mu_{\alpha\beta}^2	=	\sum_{i,j,k}  U_{\alpha j}^* U_{\beta k}  e^{-i\Delta m_{jk}^2 L/2E}  \mu_{ij} \mu_{ik},
\end{eqnarray}
where $i,j,k=1,2,3$, $\Delta m_{jk}^2 = m_j^2-m_k^2$ are the neutrino squared-mass differences, $U_{\ell i}$ is the 
leptonic mixing matrix, $E$ is the neutrino energy, $L$ is the baseline, and for simplicity we omit the electric dipole
moment contribution.

In the SM, minimally extended to include Dirac neutrino masses, NMM is suppressed by small masses of 
observable neutrinos~\cite{PDG2012} due to the left-handed nature of weak interaction. 
The diagonal and transition magnetic moments are calculated in the SM to 
be~\cite{Marciano:1977wx,Lee:1977tib,Fujikawa:1980yx,Petcov:1976ff,Pal:1981rm,Shrock:1982sc,Bilenky:1987ty,Broggini:2012df,Giunti:2008ve}
\begin{eqnarray}\label{eq:NMMlimitSM}
		\mu_{ii}^\text{SM} \approx 3.2\times10^{-20} ~\left( \frac{m_i}{0.1~\text{eV}} \right)~\mu_B
\end{eqnarray}
and
\begin{eqnarray}\label{eq:NMMlimitSMtransit1}
		\mu_{ij}^\text{SM} 	
		\approx	-4\times10^{-24} ~\left( \frac{m_i + m _j}{0.1~\text{eV}} \right)	\sum_{\ell=e,\mu,\tau} \left(\frac{m_\ell}{m_\tau}\right)^2 U_{\ell i}^*U_{\ell j}	~\mu_B,  \label{eq:NMMlimitSMtransit}
\end{eqnarray}
respectively, where $\mu_B=e/(2m_e)=5.788 \times10^{-5}$ eV\,T$^{-1}$ is the Bohr magneton. 

Currently, the strongest experimental bound on NMM is far from the SM value~\cite{Raffelt:1999gv},
\begin{eqnarray}\label{eq:NMMlimit}
	\mu_\nu < 3\times10^{-12}~\mu_B.
\end{eqnarray}
It has been obtained from the constraint on energy loss from globular cluster red giants, which can be cooled faster by the 
plasmon decays due to NMM~\cite{Bernstein:1963qh}, which delays the helium ignition. This bound can be applied to all 
diagonal and transition NMMs. 

The best present terrestrial laboratory constraints on NMM, derived in $\bar\nu_e$--$e$ elastic scattering experiments by TEXONO~\cite{Wong:2006nx},
\begin{eqnarray}\label{eq:NMMlimitTEXONO}
	\mu_{\bar\nu_e} < 7.4\times10^{-11}~\mu_B  	\qquad	(90\%~\rm{C.L.}),
\end{eqnarray}
and GEMMA~\cite{Beda:2012zz},
\begin{eqnarray}\label{eq:NMMlimitGEMMA}
	\mu_{\bar\nu_e} < 2.9\times10^{-11}~\mu_B  	\qquad	(90\%~\rm{C.L.}),
\end{eqnarray}
apply to the diagonal $\mu_{ee}$ moment, and can be translated to the transition $\mu_{e\mu}$ and $\mu_{e\tau}$ moments. However, these bounds are much weaker than the one in Eq.~\eqref{eq:NMMlimit}. 
The global fit~\cite{Grimus:2002vb,Tortola:2004vh} of NMM data from the reactor and solar neutrino experiments produces limits on the neutrino transition moments~\cite{Giunti:2008ve}
\begin{eqnarray}
	\mu_{12}, \mu_{23}, \mu_{31} < 1.8\times10^{-10}~\mu_B  	\qquad	(90\%~\rm{C.L.}).
\end{eqnarray}
NMM generically induces a radiative correction to the neutrino mass, which constrains NMM~\cite{Bell:2005kz,Bell:2006wi,Bell:2007nu}. 
In the case of diagonal NMM, which is possible only for Dirac neutrinos, the correspondent bound 
	$\mu_{\alpha\alpha} \lesssim 10^{-14}~\mu_B$
is significantly stronger than in Eq.~\eqref{eq:NMMlimit}. 
However, the transition NMM $\mu_{\alpha\beta}$, which is possible for both Dirac and Majorana neutrino types, is antisymmetric in the flavor indices, while the neutrino mass terms $m^\nu_{\alpha\beta}$ are symmetric. This may lead to suppression of the $\mu_{\alpha\beta}$ contribution to $m^\nu_{\alpha\beta}$, e.g., by the SM Yukawas, which makes the bound on NMM much weaker than in Eq.~\eqref{eq:NMMlimit}:   $\mu_{\alpha\beta} \lesssim 10^{-9}~\mu_B$~\cite{Bell:2006wi,Bell:2007nu}. 
Alternatively, Majorana neutrino masses may have spin suppression compared with NMM~\cite{Barr:1990um}.

Large NMM compared with Eqs.~\eqref{eq:NMMlimitSM} and \eqref{eq:NMMlimitSMtransit} may be generated in many theories, e.g., models with left-right symmetry~\cite{Czakon:1998rf}, scalar leptoquarks~\cite{Povarov:2007zz}, R-parity-violating 
supersymmetry~\cite{Gozdz:2012xw}, and large extra dimensions~\cite{Mohapatra:2004ce}. 
In this work we consider generation of the neutrino transition magnetic moments, using general $\nu\nu ff$ parametrization, which includes the scalar and tensor terms. 
And using the best present constraint on NMM, we get the bounds on the effective couplings.

\begin{figure}
  \centering
  \includegraphics[width=0.45\textwidth]{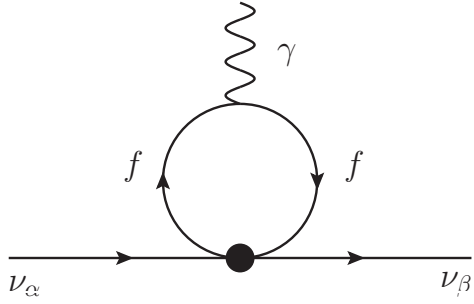}
  \caption{Effective diagram for magnetic moment of a neutrino induced by tensorial NSI, indicated by the large dot.} \label{FigNMM}
\end{figure}

We have found that among all possible $\nu_\alpha\nu_\beta ff$ interactions in Eq.~\eqref{EffLgr}, the lowest-order contribution to NMM can be generated through the one-loop diagram, shown in Fig.~\ref{FigNMM}, with the 
tensor dimension-6 operator,
\begin{eqnarray}	
	\frac{\epsilon_{\alpha\beta}^{fT}}{M^2} 		( \bar\nu_{\beta} \sigma_{\mu\nu} \nu_{\alpha}) (\bar f \sigma^{\mu\nu} f), 	
\end{eqnarray}
where in the case of Majorana neutrinos $\bar\nu_{\beta}=\bar\nu_{\beta}^c$.
In particular, interactions of neutrinos with quarks $q$ via the operator 
\begin{eqnarray}
	\frac{\epsilon_{\alpha\beta}^{q}}{M^2} (\bar\nu_\beta \sigma_{\mu\nu} \nu_\alpha) (\bar q \sigma^{\mu\nu} q), 
\end{eqnarray}
where $\epsilon_{\alpha\beta}^{q}\equiv\epsilon_{\alpha\beta}^{qT}$ is real, generate NMMs
\begin{eqnarray}\label{eq:NMMq}
	\mu_{\alpha\beta} =	\mu_{\alpha\beta}^0 - \sum_q \epsilon_{\alpha\beta}^q \frac{N_cQ_q}{\pi^2}  \frac{m_em_q}{M^2} \,  
	 \ln \left(\frac{M^2}{m_q^2}\right) \mu_B,
\end{eqnarray}
where $N_c=3$ is the number of colors, and $Q_q$ and $m_q$ are the electric charge and mass of the quark, respectively. 
Here and later $\mu_{\alpha\beta}^0$ denotes the subleading part of the NMM that is not enhanced by the large logarithm. 
We note that this formula reproduces the leading order in the exact result, which can be derived in the model with scalar LQs; 
see Ref.~\cite{Povarov:2007zz} for the exact expressions of diagonal NMMs. 

Similarly, for the interactions of neutrinos with charged leptons $\ell$,
\begin{eqnarray}
	\frac{\epsilon_{\alpha\beta}^{\ell}}{M^2} (\bar\nu_\beta \sigma_{\mu\nu} \nu_\alpha) (\bar \ell \sigma^{\mu\nu} \ell)
\end{eqnarray}
with $\epsilon_{\alpha\beta}^\ell  \equiv \epsilon_{\alpha\beta}^{\ell T}$, we have
\begin{eqnarray}\label{eq:NMMell}
	\mu_{\alpha\beta} =	\mu_{\alpha\beta}^0 +  \sum_\ell \frac{\epsilon_{\alpha\beta}^\ell}{\pi^2}  \frac{m_em_\ell}{M^2} \,   \ln \left(\frac{M^2}{m_\ell^2}\right)  \mu_B.
\end{eqnarray}
We notice that the dominant logarithmic terms, such as in Eqs.~\eqref{eq:NMMq} and \eqref{eq:NMMell}, may not contribute to NMM in certain models, e.g., in the SM, due to a mutual compensation between the relevant diagrams~\cite{Petcov:1976ff,Pal:1981rm}.

For $M=1$~TeV, using Eq.~\eqref{eq:NMMlimit} and taking one nonzero $\epsilon^f_{\alpha\beta}$ at a time, we obtain the constraints shown in Table~\ref{Tab:bounds}.
\begin{table}[htdp]
\caption{Upper bounds on the couplings $\epsilon^f_{\alpha\beta}$.}
\begin{center}
\begin{tabular}{|c|c||c|c||c|c|}
	\hline
	$|\epsilon_{\alpha\beta}^e|$	&	$3.9$	&	$|\epsilon_{\alpha\beta}^d|$	&	$0.49$	&	$|\epsilon_{\alpha\beta}^u|$	&	$0.49$ 	\\
	\hline
	$|\epsilon_{\alpha\beta}^\mu|$	&	$3.0\times 10^{-2}$	&	$|\epsilon_{\alpha\beta}^s|$	&	$3.3\times 10^{-2}$	&	$|\epsilon_{\alpha\beta}^c|$	&	$1.7\times 10^{-3}$ 	\\
	\hline
	$|\epsilon_{\alpha\beta}^\tau|$	&	$2.6\times 10^{-3}$	&	$|\epsilon_{\alpha\beta}^b|$	&	$1.2\times 10^{-3}$	&	$|\epsilon_{\alpha\beta}^t|$	&	$4.8\times 10^{-5}$ 	\\
	\hline
\end{tabular}
\end{center}
\label{Tab:bounds}
\end{table}%


Besides the limits on NMM, the neutrino-electron and neutrino-nucleus scattering~\cite{Barranco:2011wx}, as well as the matter effects in the neutrino oscillations~\cite{Bergmann:1999rz}, constrain the tensorial NSI. However, the limit on $|\epsilon^{f}_{e\beta}|$ from supernova and solar neutrino oscillations is suppressed by the small average polarization of the matter particles~\cite{Bergmann:1999rz,Nardi:2000qb}.

The tensorial contributions to the differential cross sections of $\bar\nu_e$--$e$ elastic scattering and $\bar\nu_e$--nucleus coherent scattering  can be written as~\cite{Barranco:2011wx}
\begin{eqnarray}
	\frac{d\sigma_{T}^{\nu e}}{dE_e}	=	\sum_{\beta=\mu,\tau}	\left(\epsilon^{e}_{e\beta}\right)^2	\frac{m_e}{2\pi M^4}		\left[ \left( 1-\frac{E_e}{2E_\nu} \right)^2 - \frac{m_eE_e}{4E_\nu^2} \right],
\end{eqnarray}
and
\begin{eqnarray}
	\frac{d\sigma_{T}^{\nu N}}{dE_N}	=	[|\epsilon^{u}_{e\beta}|(2Z+N) + |\epsilon^{d}_{e\beta}|(Z+2N)]^2		\frac{m_N}{2\pi M^4}		\left[ \left( 1-\frac{E_N}{2E_\nu} \right)^2 - \frac{m_NE_N}{4E_\nu^2} \right],
\end{eqnarray}
where  $m_e$ ($E_e$) and $m_N$ ($E_N$) are the masses (recoil energies) of the electron and  nucleus, respectively;  $E_\nu$ is the incident neutrino energy; and $Z$ ($N$) is the number of protons (neutrons) in the nucleus.

Using the cross section for the $\bar\nu_e$--$e$ scattering published by the TEXONO Collaboration~\cite{Deniz:2010mp,Chang:2006ug} and taking $M=1$~TeV, 
the bound $|\epsilon^{e}_{e\beta}| < 6.6$ at 90$\%$ C.L. can be obtained~\cite{Barranco:2011wx}. Using Eqs.~\eqref{eq:NMMlimitTEXONO} and \eqref{eq:NMMlimitGEMMA}, this bound can be rescaled to the GEMMA sensitivity as
\begin{eqnarray}
	|\epsilon^{e}_{e\beta}| < 2.7  	\qquad	(90\%~\rm{C.L.}).
\end{eqnarray}
The planned $\bar\nu_e$--nucleus coherent scattering experiments, e.g., part of the TEXONO low-energy neutrino program~\cite{Wong:2005vg}, can reach the sensitivity of $|\epsilon^{u,d}_{e\beta}|  < 0.2~(M/1\, \rm{TeV})^2$ at 90$\%$ C.L.~\cite{Barranco:2011wx}, which will also improve the respective bounds in Table~\ref{Tab:bounds}.

\section*{Acknowledgments}
This work was supported in part by the U.S. Department of Energy under Contract No. DE-SC0007983. The authors would like to 
thank Y.~Grossman, S.~Pakvasa and D.~Papoulias for useful comments.



\begin{thebibliography}{99}

\bibitem{Peskin:1995ev} 
See, for example, J.~Gao,  {\it et al.},
  ``The CT10 NNLO Global Analysis of QCD,''
  arXiv:1302.6246 [hep-ph];
  M.~E.~Peskin and D.~V.~Schroeder,
  An Introduction to quantum field theory
 (Addison-Wesley, Reading, PA, 1995), Sec. 17.3.

\bibitem{Artuso:2008vf} 
P.~Vilain {\it et al.}  [CHARM II Collaboration],
  Eur.\ Phys.\ J.\ C {\bf 11}, 19 (1999);
  A.~O.~Bazarko {\it et al.}  [CCFR Collaboration],
  Z.\ Phys.\ C {\bf 65}, 189 (1995);
  M.~Artuso, B.~Meadows, and A.~A.~Petrov,
  Annu.\ Rev.\ Nucl.\ Part.\ Sci.\  {\bf 58}, 249 (2008)
  [arXiv:0802.2934 [hep-ph]].

\bibitem{PDG2012}
	J. Beringer {\it et al.}  [Particle Data Group], Phys.\ Rev.\ D {\bf 86}, 010001 (2012).

\bibitem{Lee:1977tib} 
  B.~W.~Lee and R.~E.~Shrock,
  Phys.\ Rev.\ D {\bf 16}, 1444 (1977).
 
\bibitem{Grossman:1995wx} 
  Y.~Grossman,
  Phys.\ Lett.\ B {\bf 359}, 141 (1995)
  [hep-ph/9507344].

\bibitem{Bergmann:1999rz} 
  S.~Bergmann, Y.~Grossman and E.~Nardi,
  Phys.\ Rev.\ D {\bf 60}, 093008 (1999)
  [hep-ph/9903517].
  
\bibitem{Barranco:2011wx} 
  J.~Barranco, A.~Bolanos, E.~A.~Garces, O.~G.~Miranda and T.~I.~Rashba,
  Int.\ J.\ Mod.\ Phys.\ A {\bf 27}, 1250147 (2012)
  [arXiv:1108.1220 [hep-ph]].

  \bibitem{Kopp:2007ne} 
  J.~Kopp, M.~Lindner, T.~Ota and J.~Sato,
  Phys.\ Rev.\ D {\bf 77}, 013007 (2008)
  [arXiv:0708.0152 [hep-ph]].

\bibitem{Ohlsson:2012kf} 
  T.~Ohlsson,
  Rep.\  Prog.\  Phys.\  76, {\bf 044201} (2013)
  [Rept.\ Prog.\ Phys.\  {\bf 76}, 044201 (2013)]
  [arXiv:1209.2710 [hep-ph]].
    
  \bibitem{Rashed:2013dba} 
  A.~Rashed, P.~Sharma and A.~Datta,
  arXiv:1303.4332 [hep-ph].

\bibitem{Kingsley:1974kq} 
  R.~L.~Kingsley, F.~Wilczek and A.~Zee,
  Phys.\ Rev.\ D {\bf 10}, 2216 (1974).
 
\bibitem{Cho:1976um} 
  C.~F.~Cho and M.~Gourdin,
  Nucl.\ Phys.\ B {\bf 112}, 387 (1976).

\bibitem{Kayser:1979mj} 
  B.~Kayser, E.~Fischbach, S.~P.~Rosen and H.~Spivack,
  Phys.\ Rev.\ D {\bf 20}, 87 (1979).

\bibitem{Broggini:2012df} 
  C.~Broggini, C.~Giunti and A.~Studenikin,
  Adv.\ High Energy Phys.\  {\bf 2012}, 459526 (2012)
  [arXiv:1207.3980 [hep-ph]].

\bibitem{Giunti:2008ve} 
  C.~Giunti and A.~Studenikin,
  Phys.\ Atom.\ Nucl.\  {\bf 72}, 2089 (2009)
  [\phref{0812.3646}].

\bibitem{Beacom:1999wx} 
  J.~F.~Beacom and P.~Vogel,
  Phys.\ Rev.\ Lett.\  {\bf 83}, 5222 (1999)
  [hep-ph/9907383].


\bibitem{Marciano:1977wx} 
  W.~J.~Marciano and A.~I.~Sanda,
  Phys.\ Lett.\ B {\bf 67}, 303 (1977).

\bibitem{Fujikawa:1980yx} 
  K.~Fujikawa and R.~E.~Shrock,
  Phys.\ Rev.\ Lett.\  {\bf 45}, 963 (1980).

\bibitem{Petcov:1976ff} 
  S.~T.~Petcov,
  Yad.\ Fiz.\  {\bf 25}, 641 (1977)
  [Sov.\ J.\ Nucl.\ Phys.\  {\bf 25}, 340 (1977)]
  Erratum-ibid.\  {\bf 25}, 1336 (1977)
  [Erratum-ibid.\  {\bf 25}, 698 (1977)].

\bibitem{Pal:1981rm} 
  P.~B.~Pal and L.~Wolfenstein,
  Phys.\ Rev.\ D {\bf 25}, 766 (1982).

\bibitem{Shrock:1982sc} 
  R.~E.~Shrock,
  Nucl.\ Phys.\ B {\bf 206}, 359 (1982).

\bibitem{Bilenky:1987ty} 
  S.~M.~Bilenky and S.~T.~Petcov,
  Rev.\ Mod.\ Phys.\  {\bf 59}, 671 (1987)
  [Erratum-ibid.\  {\bf 61}, 169 (1989)]
  [Erratum-ibid.\  {\bf 60}, 575 (1988)].


\bibitem{Raffelt:1999gv} 
  G.~G.~Raffelt,
  Phys.\ Rept.\  {\bf 320}, 319 (1999).
  
\bibitem{Bernstein:1963qh} 
  J.~Bernstein, M.~Ruderman and G.~Feinberg,
  Phys.\ Rev.\  {\bf 132}, 1227 (1963).
  
\bibitem{Wong:2006nx} 
  H.~T.~Wong {\it et al.}  [TEXONO Collaboration],
  Phys.\ Rev.\ D {\bf 75}, 012001 (2007)
  [hep-ex/0605006].
  
\bibitem{Beda:2012zz} 
  A.~G.~Beda {\it et al.}  [GEMMA Collaboration], 
  Adv.\ High Energy Phys.\  {\bf 2012}, 350150 (2012).

\bibitem{Grimus:2002vb} 
  W.~Grimus, M.~Maltoni, T.~Schwetz, M.~A.~Tortola and J.~W.~F.~Valle,
  Nucl.\ Phys.\ B {\bf 648}, 376 (2003)
  [hep-ph/0208132].

\bibitem{Tortola:2004vh} 
  M.~A.~Tortola,
  arXiv:hep-ph/0401135.

\bibitem{Bell:2005kz} 
  N.~F.~Bell, V.~Cirigliano, M.~J.~Ramsey-Musolf, P.~Vogel and M.~B.~Wise,
  Phys.\ Rev.\ Lett.\  {\bf 95}, 151802 (2005)
  [hep-ph/0504134].

\bibitem{Bell:2006wi} 
  N.~F.~Bell, M.~Gorchtein, M.~J.~Ramsey-Musolf, P.~Vogel and P.~Wang,
  Phys.\ Lett.\ B {\bf 642}, 377 (2006)
  [hep-ph/0606248].

\bibitem{Bell:2007nu} 
  N.~F.~Bell,
  Int.\ J.\ Mod.\ Phys.\ A {\bf 22}, 4891 (2007)
  [arXiv:0707.1556 [hep-ph]].

\bibitem{Barr:1990um} 
  S.~M.~Barr, E.~M.~Freire and A.~Zee,
  Phys.\ Rev.\ Lett.\  {\bf 65}, 2626 (1990).
    
\bibitem{Czakon:1998rf} 
  M.~Czakon, J.~Gluza and M.~Zralek,
  Phys.\ Rev.\ D {\bf 59}, 013010 (1998).

\bibitem{Povarov:2007zz} 
  A.~V.~Povarov,
  Yad.\ Fiz.\  {\bf 70}, 905 (2007)
  [Phys.\ Atom.\ Nucl.\  {\bf 70}, 871 (2007)].

\bibitem{Gozdz:2012xw} 
  M.~Gozdz,
  Phys.\ Rev.\ D {\bf 85}, 055016 (2012)
  [arXiv:1201.0873 [hep-ph]].

\bibitem{Mohapatra:2004ce} 
  R.~N.~Mohapatra, S.-P.~Ng and H.-b.~Yu,
  Phys.\ Rev.\ D {\bf 70}, 057301 (2004)
  [hep-ph/0404274].

\bibitem{Nardi:2000qb} 
  E.~Nardi,
  arXiv:hep-ph/0002026.

\bibitem{Deniz:2010mp} 
  M.~Deniz {\it et al.}  [TEXONO Collaboration],
  Phys.\ Rev.\ D {\bf 82}, 033004 (2010)
  [arXiv:1006.1947 [hep-ph]].

\bibitem{Chang:2006ug} 
  H.~M.~Chang {\it et al.}  [TEXONO Collaboration],
  Phys.\ Rev.\ D {\bf 75}, 052004 (2007)
  [hep-ex/0609001].
  
\bibitem{Wong:2005vg} 
  H.~T.~Wong, H.-B.~Li, J.~Li, Q.~Yue and Z.-Y.~Zhou,
  J.\ Phys.\ Conf.\ Ser.\  {\bf 39}, 266 (2006)
  [Conf.\ Proc.\ C {\bf 060726}, 344 (2006)]
  [hep-ex/0511001].
    

\end{thebibliography}
\end{document}